\documentstyle[psfig,times]{mn2e}
\title[GRB 990413: Insight into the thermal phase evolution] {GRB
  990413: Insight into the thermal phase evolution}
  \author[Z. Bosnjak, A. Celotti, G. Ghirlanda] {\parbox[]{6.5in}
  {Z. Bosnjak$^{1,2}$, A. Celotti$^1$, G. Ghirlanda$^2$}\\ $^1$S.I.S.S.A.,
  via Beirut 2-4, I-34014 Trieste, Italy\\ $^2$INAF-Observatory of
  Brera, via Bianchi 46, I-23807 Merate (LC), Italy \\ }

\date{}

\begin{document}

\maketitle

\begin{abstract}
GRB~990413 shows a very hard spectrum (with a low energy spectral
component $F(\nu)\propto \nu^{2.49}$) which is well represented by a
black body model with characteristic temperature $\sim 70$ keV.  It
thus belongs to the subset of GRBs which might be revealing a thermal
phase.  We find that the temperature/luminosity evolution is
consistent with that found in the other ``thermal'' GRBs.  The time
resolved spectral analysis indicates the presence of a second
non--thermal component contributing (for about 1 s) up to 30 per cent of the
total flux.  Differently from the other thermal GRBs, GRB~990413 shows
significantly high level of variability and the evolution of the
thermal/non--thermal spectral components is strongly correlated with
the flux variations.  This GRB thus offers the unique opportunity to
test the standard fireball photospheric and internal shock phases and
their reciprocal influence. GRB~990413 was not selected on the basis
of its spectrum and thus hints to the possibility that this early
behavior might be more common than currently known.

\end{abstract}


\section{Introduction}

In the `standard' scenario the spectrum of gamma--ray bursts (GRBs) is
interpreted as synchrotron/inverse Compton radiation from relativistic
electrons accelerated in internal shocks within a relativistic outflow
(Rees \& M\'esz\'aros 1994). The majority of the observed spectral and
temporal GRB properties are fairly well interpreted within this
framework.  However, the nature of the prompt emission remains one 
of the most interesting unresolved problems related to GRB events.

In particular, the spectral parameter (e.g. in the Band's parametric
representation, Band et al. 1993) which is most constraining for the
emission models of GRBs is the low energy power law photon index
$\alpha$.  Synchrotron radiation from isotropic particle distributions
(Katz 1994), as well as alternative scenarios (see e.g. Lloyd \&
Petrosian 2002; M\'esz\'aros \& Rees 2000; Granot, Piran \& Sari 2000;
Ghisellini, Celotti \& Lazzati 2000) predict limits on $\alpha$ which
are consistent with at most 70 per cent of the analyzed GRB spectra
(e.g. Preece et al. 2000). Radiation mechanisms other than synchrotron
have been proposed: they involve Comptonization of a thermal
photon/particle distributions, e.g. Compton drag model (Lazzati et
al. 2000) or quasi-thermal Comptonization model (Ghisellini \& Celotti
1999; Rees \& M\'esz\'aros 2004). Furthermore, the time--resolved
spectral analysis revealed that the spectrum evolves with time
(e.g. Ford et al. 1995; Preece et al. 2000).  This might indicate an
evolution of the physical conditions of the plasma, or that various
emission processes dominate at different times and/or that the
spectral evolution is directly related to the central engine physics.

Ghirlanda, Celotti, Ghisellini (2003) [GCG03 hereafter], focusing on
the time resolved spectral analysis of a handful of extremely hard GRB
spectra, enlightened the difficulties of non-thermal models in
accounting for them and suggested that in some bursts the initial
emission phase could have a thermal character.  More recently, Ryde
(2004) [R04 hereafter] studied a few GRBs with very hard spectra,
discussing the observed emission in terms of two spectral components,
i.e.  a black body and a power-law, which showed relative variable
strengths throughout the burst durations.

In fact, in the standard GRB model (e.g. M\'esz\'aros \& Rees 2000;
Daigne \& Mochkovitch 2002) a thermal spectrum is predicted to appear
when the fireball becomes transparent (see the above references for
alternative scenarios). Initial thermal emission does not exclude the
presence of a non--thermal contribution due to the development of
internal shocks within the fireball. Therefore, we should expect to
find GRBs where both thermal and non-thermal components are present
and might evolve differently due to their different origins (e.g. Rees
\& M\'esz\'aros 2004).

In this paper we report the finding (by chance) of a new hard burst,
GRB 990413, which is particularly interesting for its spectral and
temporal properties. The emission of GRB~990413 is consistent with
thermal black body spectrum also after the initial phase and
differently from the other cases of ``thermal bursts'', its light
curve shows a high level of variability. The finest time resolved
spectral analysis reveals the presence of an underlying non-thermal
emission component and shows that the spectral evolution of the
thermal and non--thermal components is correlated with the light curve
variability.  We present the overall characteristics and the time
integrated properties of GRB 990413 in Sections 2 and 3. The analysis
and results on its spectral evolution and its correlation with the
observed light curve variability are the content of Sections 4 and 5,
respectively.

\section{GRB 990413}

GRB 990413 was detected by
BATSE\footnote{http://cossc.gsfc.nasa.gov/batse/BATSE$\_$Ctlg/index.html}.
It has a total fluence of $(6.813 \pm 0.449) \times 10^{-6}$ erg
cm$^{-2}$ in the energy range $>25$ keV. Its duration is T$_{90}=12.73
\pm 0.45$ s and its peak flux (integrated in the range 50-300 keV) is
$3.77 \pm 0.29$ phot cm$^{-2}$ s$^{-1}$, reached $\sim 2.8$ s after
the trigger. These duration and peak flux are consistent with those
typical of the long BATSE GRB population (e.g. Paciesas 1999).  The 64
ms resolution time profile integrated over the four BATSE energy
channels (i.e. for energies $>25$ keV) consists of two separate broad
pulses (Fig.~1, top) with superimposed substructures of width
$\approx10$ per cent of the main pulse width.  The broad pulses
(visually examined) span over the time intervals 0-8 s and 10-14 s
after the trigger.

\begin{figure}
%\resizebox{9.8cm}{10cm}
\psfig{file=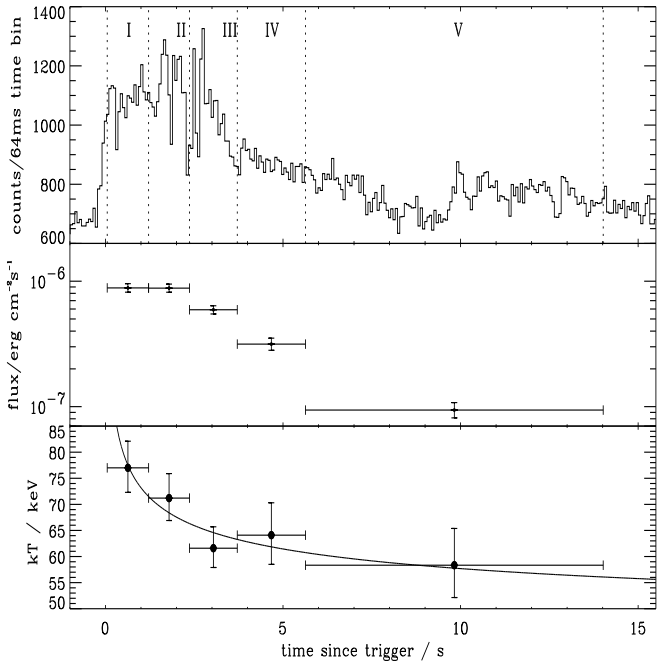,width=9.5cm,height=9cm} 
\caption{Top:
GRB 990413 light curve summed over the four BATSE energy channels ($E
>25$ keV). The five time intervals for which the time-resolved
spectral analysis was performed are indicated (see Tab.~1). Middle:
Best fit BB model flux integrated in the energy fit range.  Bottom:
black body temperature evolution and the fit with a kT $\propto$
~t$^{-1/4}$ function (solid line).}
\end{figure}

\section{Time integrated spectrum}

The spectral analysis was performed on the High Energy Resolution
Burst (HERB)
data\footnote{ftp://cossc.gsfc.nasa.gov/compton/data/batse/trigger},
which are typically composed by 120 energy channel spectra distributed
in the range $\sim30 - 1500$ keV, and have a rate--dependent
accumulation time. The spectra were selected from the most illuminated
Large Area Detector (LAD \#5) in order to have the highest
signal-to-noise ratio (S/N) (e.g. Band et al. 1993).  The background
spectrum was estimated as the average spectrum over two time intervals
before and after the burst trigger (2070 s and 19 s, respectively).

The count spectrum, accumulated over the GRB total duration of 14 s
after the burst trigger, was fitted with the Band model,
i.e. parametrized by a smoothly connected double power-law with low
and high energy spectral indices $\alpha$ and $\beta$, and a break
energy $E_{\rm b}$ (Band et al. 1993, Preece et al. 2000). This model
fit resulted in $\chi^2_{\rm r}=1.27$ for 95 d.o.f., with an extremely
hard low energy spectral slope, i.e. $\alpha=1.52^{+0.32}_{-0.30}$ and
an unconstrained $\beta$ due to the low signal above $\sim 600$ keV.
We adopted therefore the COMP model, i.e. a power law with a high
energy exponential cutoff
$N(E)=K(E/1\textrm{keV})^{\alpha}e^{-E/E_{\rm b}}$, to fit the
spectrum \footnote{The fit energy range was 43 - 1787 keV; the low
energy end is not the standard 30 keV as we had a large energy channel
width (8 keV) at 32 keV after the channel grouping to obtain a higher
S/N.} (Fig.~2).  With this model the reduced $\chi^2_{\rm r}=1.25$ for
96 d.o.f. was formally obtained; the COMP model best fit also
indicated a very hard low energy spectral component, with photon
spectral index $\alpha=1.49_{-0.27}^{+0.34}$, and an e-folding energy
$E_{\rm b}=75.2_{-5.0}^{+5.1}$ keV, which corresponds to a peak energy
$E_{\rm peak}=E_{\rm b}(\alpha+2) \approx260$ keV in the $E\, F_{\rm
E}$ spectrum.

\begin{figure}
\psfig{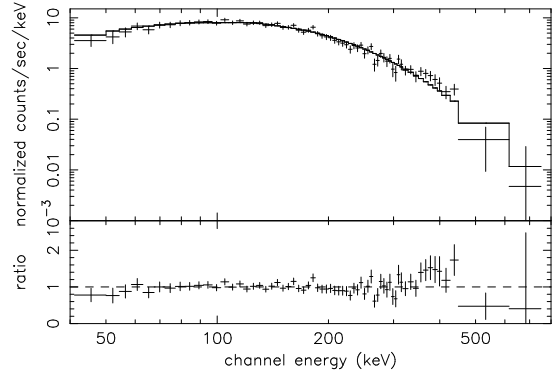}
\caption{COMP model fit to the time integrated (14 s) count
spectrum. The solid line represents the best fit model with
$\alpha=+1.49^{+0.34}_{-0.27}$ and $E_{b}=75.2\pm5.0$. The lower panel
shows the model-to-data ratio.}
\end{figure}
As a possible test of the consistency with the synchrotron model, we
also fitted the spectrum with a COMP model with frozen low energy
spectral photon index $\alpha=-2/3$, i.e. the limiting value of the
optically thin synchrotron emission (Tavani 1996). The fit resulted in
a reduced $\chi^2_{\rm r} \sim 4$ for 97 d.o.f., rejecting the model
at 99.9 per cent confidence level.

Given the extreme hardness of the low energy spectral component, we
fitted the time integrated spectrum of GRB~990413 with a Black Body
model (BB). The best fit ($\chi^2_{\rm r}$=1.27 for 97 d.o.f.)
corresponds to a temperature $kT=67.9^{+2.3}_{-2.3}$ keV, i.e. a peak
energy $E_{\rm peak}\sim$196 keV.  This value is consistent with the
few hundred keV peak energies found in the other 10 GRBs found (so
far) to be consistent with thermal spectra (GCG03 and R04).  We also
note that, although the GRB spectrum is well fitted by a BB model, its
low energy slope is slightly steeper than the Rayleigh--Jeans
extrapolation in the low energy domain (i.e. $\alpha=1$), as already
indicated by the best COMP fit which gives $\alpha\sim1.5$.

Following the procedure described in GCG03 we also tested the
robustness of the spectral fits. To this aim: 1) we performed the fits
reducing the energy range at the high energy end to assess whether the
characteristic exponential cutoff energy of the COMP model might
determine a hard low energy spectral component and we found that
$\alpha$ did not significantly change within the confidence interval
centered around the above best fit value; 2) we examined the (null)
influence of different background spectra by choosing different time
intervals to compute the average background; 3) we used the data from
the second ranked LAD (\# 7). The spectral fit resulted in a still
positive, but less steep $\alpha$=0.29$^{+0.06}_{-0.28}$. Because of
the lower S/N ratio, in this case $\chi_r^2$=1.61 for 97 d.o.f.; 4) as
a check of the possible dependence on the detector response, we used
the detector response matrix (DRM) corresponding to another GRB to fit
the spectrum of GRB 990413. To this purpose we searched for a GRB with
approximately the same angle of incidence and sky position and adopted
its DRM to fit the spectrum of GRB 990413.  The fit with the COMP
model gives $\alpha= 1.11_{-0.19}^{+0.27}$, with $\chi_{\rm r}^2$=1.26
for 97 d.o.f. We conclude that the result that GRB 990413 has an
extremely hard low energy spectrum, consistent with a BB model, is
indeed robust.

\section{Time resolved spectral analysis}

As well known the GRB spectrum typically evolves in time with no
unique behavior. The spectral evolution is evident in all of the
three spectral parameters, i.e. low and high energy spectral slopes
and peak energy. For instance, the asymptotic $\alpha$ evolves in 58
per cent  of the bursts (Crider et al 1997). Also the GRBs with a thermal
spectrum show evidence of a spectral evolution which is well described
by the best fit temperature. GCG03 found that in their 5 GRBs the BB
temperature, after an initial constant phase (of few tenths of a
second), evolves as $\sim t^{-1/4}$. A slightly steeper decay, $\sim
t^{-2/3}$, was found in other 5 `thermal' bursts by R04. The luminosity
evolution instead does not show a unique behavior.

We performed a time resolved spectral analysis using HERB data (see
Table~1), covering the time interval $0.05-14$ s after the trigger
time, for a total of 5 spectra with similar integration time ($\sim$
1.2 s each, except the last one).  We fitted them with both the COMP
and the BB model. The results are shown in Table~1. The reduced
$\chi^{2}_{\rm r}$ is acceptable in almost all cases (a possibly larger 
value is found only for the last
spectrum which is indeed characterized by a lower S/N). The typical
values $\alpha$ for the COMP model are similar to that found from the
analysis of the time integrated spectrum (i.e. $\alpha\sim 1.5$ -
Sec.~2). Only the second spectrum shows a slightly softer low energy
spectral component (see Sec.~3 for a possible explanation). The peak
spectral energy $E_{\rm peak}$ decreases from an initial value of
$\sim 296$ keV to $\sim 222$ keV in the late time spectrum, indicating
a typical hard-to-soft evolution.

In Fig.~1 (middle and bottom panels) we report the evolution of the
spectral parameters of the BB fit (see $kT_{\rm BB}$ in Tab. 1). The
BB flux (integrated in the fit energy range) is nearly constant for
the initial 2.36 s and then decreases approximately as $\propto t^{-1}$. The
temperature evolution is instead well described by $kT \propto
t^{-1/4}$ (solid line in the bottom panel of Fig.~1). This behavior
fully agrees with what found by GCG03 for their set of `thermal' GRBs.

\begin{table}
{\small \begin{tabular}{cccccc} \hline\noalign{\smallskip}
%$\Delta E_{\rm fit}$ &
$\Delta t_{\rm fit}$ [s] & model & $\alpha$ & $E_{\rm b}$ [keV] & kT$_{\rm BB}$ [keV]& $\chi_{\rm r}^2$/d.o.f.
\\
\noalign{\smallskip}
% s &   &  & keV & keV &  \\
\noalign{\smallskip}
\hline
\hline
\noalign{\smallskip}
 0.05--1.21 & comp & $1.50^{+0.47}_{-0.53}$ &$84.7_{-8.3}^{+15.7}$ & &1.17/88
\\
          & bb  &  & & $77.0_{-4.7}^{+5.1}$ & 1.17/89 \\
\noalign{\smallskip}
 1.21--2.36 &comp  & $0.81^{+0.42}_{-0.50}$ & $109.1_{-21.6}^{+43.6}$ & 
&1.04/73 \\
         & bb & & & $71.2_{-4.3}^{+4.7}$ & 1.10/74 \\
\noalign{\smallskip}
 2.36--3.71 & comp & $1.51^{+0.74}_{-0.56}$ &$67.8_{-8.4}^{+19.3}$ & &0.85/87
\\
      & bb  &  & & $61.6_{-3.7}^{+4.1}$ & 0.85/88 \\
\noalign{\smallskip}
 3.71--5.63 & comp & $1.53^{+0.58}_{-0.68}$ &$69.7_{-10.3}^{+24.4}$ & &1.04/99
\\
         & bb  &  & & $64.1_{-5.6}^{+6.2}$ & 1.03/100 \\
\noalign{\smallskip}
 5.63--14.01 & comp & $1.55^{+0.47}_{-0.17}$ &$62.5_{-5.8}^{+8.2}$ & &1.36/104
\\
          & bb  &  & & $58.3_{-6.2}^{+7.0}$ & 1.36/105 \\
\noalign{\smallskip}
\hline
\end{tabular}}
\caption{Results of the time resolved spectral analysis of GRB~990413.
The COMP and BB best fit spectral parameters are reported for each
time resolved spectrum. Errors on the parameters correspond to the
90 per cent confidence level. $\Delta t_{\rm fit}$ is the time interval over
which each spectrum is accumulated. The energy fit range is
approximately 40-1800 keV for all cases, except for the second
spectrum which was analyzed on the reduced energy range 40-900 keV,
due to the extremely low count number in the high energy channels.}
\end{table}

The 10 BATSE GRBs whose spectra are consistent with a thermal model
present a smooth light curve (see GCG03 and R04, Fig.~3 and Fig.~1
respectively). GRB~990413 instead is quite variable, with a light
curve characterized by rapid dips (see Fig.~4) during which the count
rate is reduced by a factor $\sim$ 3 with respect to its value at the
peaks. In view of understanding what is happening to the thermal
component during such rapid drops in intensity we examined the
spectral evolution on (even shorter) timescales, comparable with the
light curve variability.

However, the LAD data have a maximum (S/N limited) time resolution
$>1$ s. For this reason we analyzed the Medium Energy Resolution (MER)
data that have a minimum accumulation time of 16 ms, despite of a
lower energy resolution (16-energy channels) compared to the HERB
spectra.  We integrated the spectra in time bins tracking the `peaks'
and 'dips' of the light curve; the selected time intervals are shown
in Fig. ~3, top panel, and reported in the first column of Table~2.

For all the spectra, in 15 selected time intervals, we first tested
the BB model, and found that it does still give acceptable fits,
except for those corresponding to the 4 `dips' identified in the light
curve. In these latter cases we found that the spectrum is best fitted
by a single power-law (PL) model. This appears to suggest the presence
of a second non--thermal spectral component.  Indeed the known GRBs
with evidence of thermal spectra exhibit a smooth, continuous
evolution in time, from an initial thermal phase to a non-thermal one
(GCG03) and a double model fit, i.e. BB+PL, has been used by R04.

We thus systematically analyzed all the 15 spectra with both the BB
and the PL model (separately and jointly) and report the results in
Tab.~2 and graphically in Fig.~3.  The four `dip' spectra (integrated
over 48, 96 and 80 ms for the last two) are best fit with a PL model
without a thermal contribution. The power-law photon spectral index is
very similar in all the 4 dips, i.e. $\alpha\sim -1.6$.  If this
corresponds to the photon index $\alpha$ typical of non--thermal
spectra found in long bright GRBs (e.g. Preece et al.  2000), the peak
energy should be $>1500$ keV i.e.  the upper energy limit of the fit
window. We note that this spectral slope is consistent with those
expected for synchrotron emission from a cooling population of
relativistic electrons (such as those accelerated in internal
shocks). Interestingly, we also find that the spectrum corresponding
to the peak preceding the first dip and between the first and second
one is better fit by a BB+PL model.

\begin{figure}
%\resizebox{9.8cm}{!}
\psfig{file=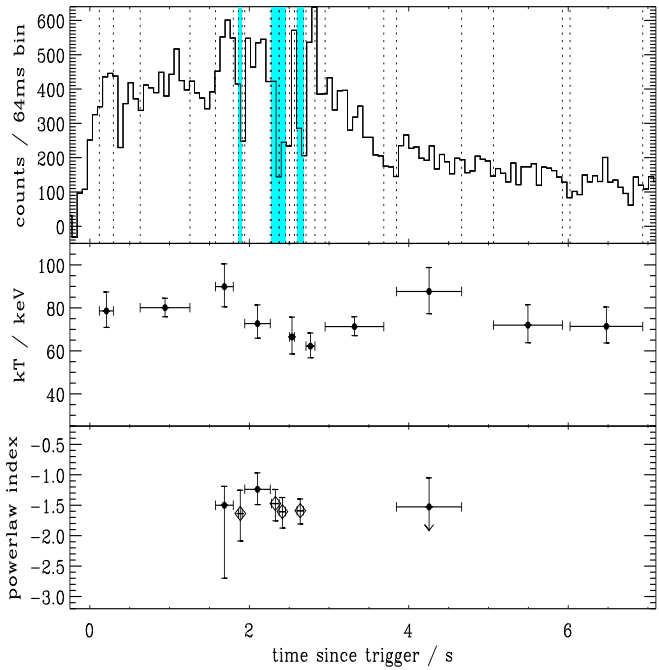,width=9.5cm,height=9cm}
\caption{Results of the spectral analysis of GRB~990413 with high
temporal resolution. For clarity only the first pulse of the light
curve and its decay are shown; the second pulse is integrated in only
one spectrum in the interval 9.9--11.3 s (due to the lower S/N) as
reported in Table~2.  Top panel: count rate integrated over the four
BATSE energy channels.  The shadowed regions highlight the `dips' of
the light curve where a non--thermal contribution dominates. Middle
and bottom panels: temporal evolution of the spectral parameters.
Diamonds indicate the parameters corresponding to the dips.  When both
$kT$ of the BB fit and the PL index are reported for the same time
interval, a two-component model gave the best fit.}
\end{figure}

In the spectra where both components (BB and PL)
contribute to the total photon number, we find that the BB
component dominates: it  comprises $>$80 per cent of the total
emission during the first peak and $>$60 per cent during the second one
(see Fig.~4).

\begin{figure}
\psfig{file=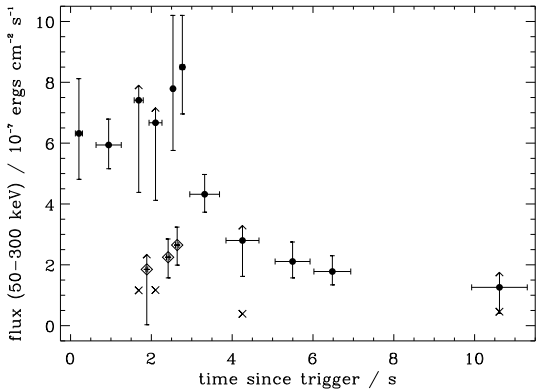,width=8cm}
\caption{Flux in the 50-300 keV energy band; circles refer to the flux
integrated in the time intervals in which the `peaks' in the light
curve occur and diamonds correspond to flux during `dips'. The flux in
the interval 2.28-2.37 s is not reported because the associated fit
normalization is unconstrained.  The crosses
indicate the contribution of the power law component to the total flux
when the best fit comprises a black body superimposed to a power law
component.}
\end{figure}

\begin{table}
{\small \begin{tabular}{ccccc} \hline\noalign{\smallskip}
%$\Delta E_{\rm fit}$ &
$\Delta ~ t_{\rm fit}$ [s] & $p$ & kT$_{\rm BB}$ [keV] & $\chi_{\rm r}^2$ / d.o.f.&
peak/dip \\
\noalign{\smallskip}
\hline
\hline
\noalign{\smallskip}
 0.12 -- 0.29 & -- &$78.6_{-7.7}^{+8.8}$ &1.51/11 & peak \\
\noalign{\smallskip}
 0.63 -- 1.25 & --& $80.1_{-4.2}^{+3.4}$   &1.73/11 &peak \\
\noalign{\smallskip}
 1.57 -- 1.78  & $-1.5_{-1.2}^{+0.3}$ &$89.9_{-9.4}^{+10.6}$ &1.31/9 &peak \\
\noalign{\smallskip}
 1.86 -- 1.91  & $-1.6_{-0.4}^{+0.4}$ &--&0.46/10& dip \\
\noalign{\smallskip}
 1.94 -- 2.26  & $-1.2_{-0.3}^{+0.3}$ &$72.9_{-6.8}^{+8.5}$  &0.78/9 &peak \\
\noalign{\smallskip}
 2.28 -- 2.37  & $-1.5_{-0.3}^{+0.2}$ &--  &1.26/9 &dip \\
\noalign{\smallskip}
 2.37 -- 2.45 & $-1.6_{-0.3}^{+0.2}$ & -- &0.77/11 &dip \\
\noalign{\smallskip}
 2.50 -- 2.57  & -- &$66.5_{-7.9}^{+9.2}$  &0.99/11 & peak \\
\noalign{\smallskip}
 2.59 -- 2.68  & $-1.6_{-0.2}^{+0.2}$  & --  & 1.45/11 &dip \\
\noalign{\smallskip}
 2.71 -- 2.82  & -- &$62.2_{-5.4}^{+6.1}$  &1.43/11 &peak \\
\noalign{\smallskip}
 2.95 -- 3.69  & -- &$71.3_{-4.2}^{+4.6}$  &1.57/11 &peak \\
\noalign{\smallskip}
 3.85 -- 4.66  & $-1.5_{-3.2}^{+0.5}$ &$87.6_{-10.3}^{+11.1}$  &0.29/9&peak \\
\noalign{\smallskip}
 5.06 -- 5.92  & -- &$72.0_{-8.2}^{+9.5}$  &1.78/11 &peak \\
\noalign{\smallskip}
 6.02 -- 6.93  & -- &$71.4_{-7.7}^{+8.9}$  &1.37/9 &peak \\
\noalign{\smallskip}
 9.93 -- 11.30  & $-1.3_{-0.4}^{+0.4}$ &$54.2_{-11.1}^{+17.7}$  &0.33/9 &peak
\\
\noalign{\smallskip}
\hline
\end{tabular}}
\caption{Results of the spectral analysis with high temporal
resolution for a BB and PL models fitted to the MER data of
GRB~990413. When only $kT$ is reported, the spectrum was best fitted
by a BB component alone. $\Delta t_{\rm fit}$ is the time interval
over which the spectra were integrated.  The energy fit range is
approximately 36-1548 keV for all spectra. $p$ is the single
power law photon spectral index defined as $N(E)\propto E^{p}$.}
\end{table}

\section{Variability estimate}

In the previous section it has been shown that the spectral components
(thermal and non--thermal) present in GRB~990413 appear to be
correlated to the variability of its light curve. In this respect
GRB~990413 presents a new, unique phenomenology. The light curves of
the other 10 `thermal' bursts are qualitatively smoother than that of
GRB~990413 which is instead similar to the majority of GRBs. In this
section we quantitatively assess these statements by comparing the
variability of GRB~990413 with that of typical non--thermal bursts and
with the small sample of the thermal GRBs.

In order to evaluate the observed variability we applied
the method proposed by Fenimore $\&$ Ramirez-Ruiz (2000). We
computed the variability as
\begin{equation}
V=\frac{1}{N} \sum_i \frac{( c_i - < c >_{\rm w})^2 -(b_i + c_i)}{c_{\rm p}^2},
\end{equation}
where c$_i$ are the source photon counts (at 64 ms resolution); $N$ is
the number of time bins with a signal $> 5\sigma$ above the
background; c$_{\rm p}$ is the number of peak counts used for
normalization; $<$c$>_{\rm w}$ are the counts smoothed with the boxcar
window of width $w$. The choice of the time interval for smoothing was
determined by the average (observed) thermal phase duration, i.e.  $w$
is comparable or shorter than the time interval in which the thermal
spectra were observed (clearly longer time scales would not reflect
the fast fluctuations that we want to sample).  We performed several
tests and the optimal choice was found to be $w$ = 3 s. As the
observed time history should be corrected for the cosmological time
dilation, we also verified how the result would change if a burst was
placed at a typical redshift $z$=1: we found - applying the correction
as defined in Fenimore $\&$ Ramirez-Ruiz (2000) - that the variability
changes by less than 10 per cent (the major effect is produced by
variations in the smoothing time window).  

As the uncertainties on the measured variability $V$ (see Reichart et
al. 2001 for the detailed procedure to estimate them) strongly depend
on the brightness of the event, for the comparison between GRB~990413
and a larger set of bursts we computed $V$ for a sample of 217
BATSE GRBs of comparable durations (T$_{90}>$10 s) and peak fluxes
(3.7 phot s$^{-1}$ cm$^{-2} < $ P$_{\rm 64ms} <$ 30.7 phot s$^{-1}$
cm$^{-2}$). This latter range includes the fluxes measured for
GRB~990413 and the bursts analyzed in GCG03. The results are shown in
Figure 5: the variablity of GRB~990413 is clearly similar to the
majority of the GRBs in the sample, while it exceeds (by a factor
$\sim$ 10 in $V$) the rest of the `thermal' bursts. We verified that
this does not depend on the choice of $w$, and found that the
variability of GRB~990413 increases proportionally to the length of
the smoothing timescale with respect to the other hard-thermal
bursts. The 5 GRBs analyzed by R04 have lower peak fluxes and for the
given smoothing time scale ($w$=3 s) their variability ranges
from $\sim 7\times 10^{-5}$ to $4\times 10^{-3}$ (although with large
uncertainties due to the low S/N), i.e. lower than that of
GRB~990413.

\begin{figure}
\psfig{file=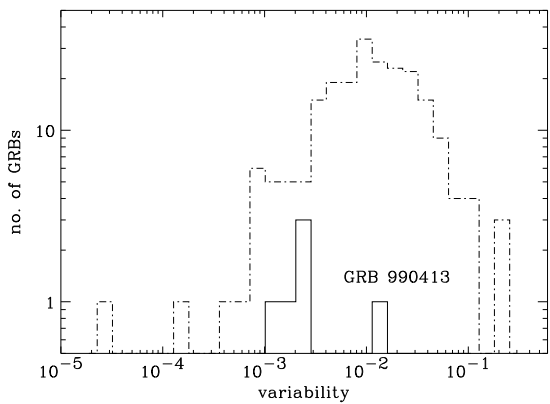,width=8.2cm}
\caption{Variability measure for a smoothing time window $w$=3 s; the
dot-dashed histogram represents the sample of 217 GRBs selected within
peak flux and duration ranges comprising the values of the bright
thermal GRBs (GCG03) and of GRB~990413.  GRB~990413 is indicated on
the plot.}
\end{figure}

\section{Summary and discussion}

We presented the time integrated and the time resolved spectral
analysis of the long duration (14 s) double peaked BATSE burst
GRB~990413 that was found to have an extremely hard spectrum.  Our
analysis indicates that (i) the average spectrum is consistent with
thermal emission with typical temperature $kT\sim70$ keV, comparable
to the other 10 'thermal' bursts, and (ii) the spectral evolution of
this component is consistent with what found for the thermal GRBs, i.e
$kT \propto t^{-1/4}$.

However, we find evidence that a second non--thermal component
contributes to the spectrum during the main pulse, in correspondence
with the highly variable phase.  A combined model (thermal +
non--thermal) has been fitted to the time resolved spectra, showing
that: (i) the non--thermal component is present only for about 1 s
during the long (14 s) burst, (ii)  its slope does not evolve in time
and indicates a hard spectrum and (iii) it dominates the
flux only in correspondence of the minima (`dips') of the GRB light
curve, while the thermal component dominates for all of the rest of
the burst duration.

In the standard GRB fireball model thermal emission is expected from
the expanding fireball when it becomes transparent and typically
before the non-thermal particle acceleration in internal shocks
occurs.  This scenario for the two emission phases can account for an
initial thermal spectrum evolving into a non--thermal one, as observed
in the 10 GRBs studied by GCG03 and R04.  GRB~990413, besides the
hard-thermal component, revealed a non--thermal contribution still
during the (possibly) photospheric phase.  Moreover, this component is
associated with the rapid temporal variability ($V$=0.016) that is a
result, in the internal shock scenario, of the velocity distribution
of the colliding shells within the relativistic outflow.  Therefore,
any model aimed at accounting for the spectral properties of
GRB~990413 needs to allow for the possibility of developing internal
shocks (or other forms of dissipation) inside the photosphere
(e.g. Daigne \& Mochkovitch 2002; Rees \& Meszaros 2004), and has to
reproduce consistently the evolution of their relative contributions.

The presence of rapid dips during a thermal dominated phase is also
interesting.  Among the possible causes two appear quite plausible:
(i) there might be inhomogeneities in the fireball optical depth (at
the time of transparency); (ii) the dips directly trace the central
engine intermittence in energy injection and launching of the fireball
(the dips duration still largely exceed the typical timescales
associated with the fireball formation).

While it is not easy to estimate the percentage of GRBs which may show
a detectable thermal component, the fact that GRB~990413 was found
independently of its spectral properties may indicate that this
phenomenology is more common than known so far.  It should be noticed
that the at the brightness level of GRB 990413, the consistency of the
spectrum with such a black-body like component is not due to low
signal-to-noise ratio.

A detailed numerical calculation of the `transition' to transparency
and the formation of internal shocks might clarify the relative roles
of the two different radiative regimes and provide information on the
fireball physical parameters (Lorentz factor distribution of the
shells, baryon loading).

\section{Acknowledgments}

It is a pleasure to thank Enrico Ramirez-Ruiz for useful discussions
and suggestions. We also thank the referee for constructive
comments. This research was supported in part by the National Science
Foundation under Grant No. PHY99-07949; the KITP (Santa Barbara) is
thanked for kind hospitality (AC).  The Italian MIUR and INAF are acknowledged 
for financial support.

\end{document}